\documentclass[11pt]{article}
\usepackage{amsmath,amsfonts,amssymb,color,graphicx,latexsym,theorem,mathrsfs
}

\newcommand{\ma}[1]{\mbox{$\mathcal{#1}$}}
\newcommand{\mas}[1]{\mbox{$\mathscr{#1}$}}

\newcommand{\D}{{\rm d}}
\newcommand{\ti}{\tilde}
\newcommand{\we}{\wedge}

\begin{document}

\begin{titlepage}

\begin{flushright}
{
\small IFUM-1030-FT\\
\today
}
\end{flushright}
\vspace{1cm}

\begin{center}
%---------- title ----------%
{\huge  Positive mass theorem in extended supergravities}
\end{center}
\vspace{.5cm}

\begin{center}
%---------- author ----------%
Masato Nozawa\footnote{{\tt masato.nozawa`at'mi.infn.it}}
and Tetsuya Shiromizu\footnote{{\tt shiromizu`at'math.nagoya-u.ac.jp}}

\vspace{.4cm}

%---------- affiliation ----------%
{\it ${}^1$Dipartimento di Fisica, Universit\`a di Milano,
and INFN, Sezione di Milano, \\
Via Celoria 16, 20133 Milano, Italia \\
${}^2$Department of Mathematics, and Kobayashi-Maskawa Institute, \\
Nagoya University, Nagoya 464-8602, Japan 
}

\end{center}

\vspace{.5cm}

%---------- abstract ----------%
\begin{abstract}
Following the Witten-Nester formalism, we present a useful prescription using Weyl spinors towards the positivity of mass.  
As a generalization of arXiv:1310.1663, we show that some ``positivity conditions'' must 
be imposed upon the gauge connections appearing in the supercovariant derivative acting on spinors. 
A complete classification of the connection fulfilling the positivity conditions is given. 
It turns out that these positivity conditions are indeed satisfied for a number of  
extended supergravity theories. 
It is shown that the positivity property holds for the Einstein-complex scalar system, provided that
the target space is Hodge-K\"ahler and the potential is expressed in terms of the superpotential. 
In the Einstein-Maxwell-dilaton theory with a dilaton potential, the dilaton coupling function and the 
superpotential are fixed by the positive mass property. 
We also explore the $N=8$ gauged supergravity and demonstrate that the 
positivity of the mass holds independently of the gaugings and the deformation parameters. 
\end{abstract}

\vspace{.5cm}

\setcounter{footnote}{0}

\end{titlepage}

\tableofcontents

\section{Introduction}

The positive mass theorem~\cite{Schon:1979rg,Schoen} is one of the major achievements 
in mathematical theory of general relativity.  
If the positivity property of the mass fails to be valid, 
the vacuum Minkowski spacetime which obviously has a vanishing mass
possibly decays into configurations with lower energy, 
and a dynamical ``chasing instability'' is unavoidable due to the weak equivalence principle~\cite{Bondi:1957zz}. 
The positive mass theorem therefore forbids these 
undesirable phenomena and accounts for the stability of the 
lowest energy states. 

Since the first proof given by Schoen-Yau~\cite{Schon:1979rg,Schoen},  
various attempts have been done towards the generalization. 
This subject is stimulated not only by a purely mathematical interest. 
First of all, the proof of Schoen-Yau  cannot be applied to 
$D\ge 9$ dimensions, since the smoothness of 
the deformation of the $n$-dimensional minimal surface $S_n$ is guaranteed 
for $n\le 6$.  The proof based on the inverse mean curvature flow~\cite{JW,HI} provides 
a physically clear interpretation. However, it has been successful only in $D=4$
since the Gauss-Bonnet theorem over the two-surface was explicitly used therein.  
Furthermore, both of these methods work only in the spacetimes that 
tend asymptotically to the Minkowski spacetime. 
Compared to these proofs, 
a remarkably simple and elegant proof was given by Witten~\cite{Witten}, later refined by
Nester~\cite{Nester}. A distinguished feature of their proof is the use of a spinor field.  
The bilinear vector built out of the spinor field used in their proof plays   
the role of the infinitesimal generator of the asymptotic symmetry. Although the use of spinor 
imposes a mild restriction upon the spacetime topology,\footnote{
The condition that the spacetime admits the spin structure amounts to requiring 
that the second Stiefel-Whitney class should vanish. 
Some five-dimensional asymptotically flat soliton solutions 
found in refs.~\cite{Compere:2009iy,Gibbons:2013tqa}  violate the 
mass bound proven by the spinorial method in~\cite{Gibbons:1993xt}, since they fail to possess the spin structure.  
It is an interesting but a challenging task to derive the 
lower bound of five-dimensional Arnowitt-Deser-Misner (ADM) mass 
in Einstein-Maxwell-Chern-Simons gravity
without assuming the spin structure. 
} this proof is sometimes more powerful since it is 
able to give a strictly positive bound on the mass, 
rather than a simple positivity thereof. 
Moreover, the Witten-Nester approach has additional advantages that it works in arbitrary dimensions, 
it  is applicable also for asymptotically anti-de Sitter (AdS) spacetimes and it does not necessarily 
require the dominant energy condition. 
Another utility of using spinors is that it possess an intimate relationship 
to supergravity theories~\cite{Deser:1977hu,Horowitz:1982mu,Hull:1983ap,Deser:1983rn}.

Recently, a number of widespread revival interests in extended supergravities have been growing 
from the viewpoint of string theory and AdS/CFT correspondence. 
Among others, the supersymmetric solutions in supergravity have played a 
central role in their theoretical development. Since supersymmetric solutions 
belong to the short multiplets, they are essentially nonperturbative objects, hence they 
usually evade instabilities. 
They are characterized by the existence of Killing spinors 
obeying the 1st-order differential equations~\cite{Maeda:2011sh}. 
Similar to the Bogomol'nyi-Prasad-Sommerfield (BPS) states in solitons, they are 
often identified as states saturating a certain kind of inequality between 
conserved quantities implied by the  positive mass theorem. 
Note that this is not obvious since the quantities in the superalgebra
are associated with the invariance of the background spacetime, i.e., they do not correspond to the 
conserved quantities in the general curved spacetime which approaches asymptotically to that background. 

Thus far, various supergravity theories have been shown to admit the 
BPS bound~\cite{Gibbons:1982fy,Gibbons:1982jg,Gibbons:1983aq,Boucher:1984yx,Townsend:1984iu,Gibbons:1994vm,Kostelecky:1995ei,Shiromizu:1999xm,Rogatko:2002qe}, in terms of globally conserved quantities. 
It should be worth commenting that the converse statement is not always true, namely, 
the theory admitting the BPS-type inequality does not necessarily have a supergravity origin.  
For example,  the Einstein-Maxwell-dilaton theory admits the BPS-type inequality~\cite{Gibbons:1993xt}. 
It was realized, however,  that the 1st-order BPS equation for the saturation of inequality is incompatible with the 
equations of motion except for the particular values of the coupling constant~\cite{Nozawa:2010rf}. 
This implies that it is not always possible to embed the theory admitting 
BPS-type inequality into supergravity. 

At the current moment, 
it is also  less obvious which theories admit the BPS-type inequality, when 
the supergravity embedding is unknown. 
In our previous paper~\cite{NS}, we tackled this problem 
pursuant to the Witten-Nester argument, and 
found that a certain condition should be imposed  toward the positivity bound
upon the connection in the supercovariant derivative 
acting on a Dirac spinor.  
By virtue of this condition, we were able to construct the first instance of noncanonical scalar-field
system admitting the BPS-type inequality~\cite{NS} (see also \cite{Elder:2014fea}). 
In the current article, we 
generalize the argument in~\cite{NS} and reformulate 
the ``positivity conditions'' in terms of Weyl spinors. 
We also provide a proof for the classification of connections 
satisfying the positivity conditions. 
This would make it clear the relationship to the four-dimensional 
extended supergravity theories.  In $N=1$ supergravity, it has been widely 
known that the theory admits the positive mass~\cite{Deser:1977hu}.
The $N>1$ case is less clear since extended supergravities do not always 
have an $N=1$ description except for  the consistent truncation. 
The purpose of the present paper  
is to examine the positivity property of various theories inspired by extended supergravities.  
Using the positivity conditions, we  
resolve some issues about the BPS-type inequality in extended supergravities, and  
demonstrate that the positivity property is indeed true for wider theories than formerly considered.

The plan of the present paper is as follows. In the next section, 
we formulate the Witten-Nester method in terms of Weyl spinors 
and address the positivity of the Witten-Nester energy.  We find that the gauge connections 
appearing in the supercovariant derivatives should satisfy 
the ``positivity conditions.'' This is a generalization of our previous work~\cite{NS}.
A classification of the connections satisfying the positivity conditions is given in 
appendix, where it is shown that the possible connections take the same form as 
those appearing in extended supergravity, 
provided we impose an additional condition that the bilinear vector is a Killing field for the BPS geometry.   
In section~\ref{ex}, we apply this formalism to various theories inspired by supergravity. 
We resolve some problems in the literature and find that the positivity of energy 
holds in in much broader class of theories than previously studied. 
In particular the maximal gauged  supergravity turns out to admit the mass positivity, 
independent of the gaugings and symplectic frames. 
The final conclusion with some future prospective works is 
described in section~\ref{summary}.

Our conventions for the metric is taken to be mostly plus sign. 
$\mu, \nu,...$ refer to the spacetime indices, whereas $a,b,...$ to the 
frame indices. We adopt the units $c=8\pi G=1$ throughout the paper.

\section{Positive mass theorem \`a la Witten-Nester}
\label{sec:PMT}

In our previous paper~\cite{NS}, we derived a minimal condition toward the positive mass 
for the  gauge connection in the supercovariant derivative acting on a Dirac spinor. 
This condition provides a universally simple formula and 
is able to easily recover all of the previous positive mass results. 
In the present paper we are interested in theories inspired by extended 
supergravities. Hence it turns out to be more advantageous to generalize the 
analysis~\cite{NS} in terms of Weyl spinors. We shall restrict exclusively to
four dimensions for simplicity, although the higher (even) dimensional 
extension is straightforward. 
We will work in mostly plus metric signature and the Clifford algebra reads 
$\{\gamma_a, \gamma_b \}=2\eta_{ab}=2{\rm diag}(-1,1,1,1)_{ab}$.  
Taking the orientation as $\epsilon_{0123}=1$, 
the chiral matrix is defined by $\gamma_5=-(i/4!) \epsilon_{abcd}\gamma^{abcd}=i \gamma_{0123}$ 
with $\gamma_5^2=1$. 
The (anti-)self dual part $H^\pm$ of the 2-form $H_{\mu\nu}$ is 
$H^\pm =\frac 12 (H \mp i \star H)$, satisfying $\star H ^\pm =\pm i H^\pm$.

We denote the set of Weyl spinors 
in four dimensions by $\epsilon_i $ ($i=1,...,N$). 
We take these spinors to have a negative chirality 
$\gamma_5 \epsilon _i=-\epsilon_i$. 
If we define the Dirac conjugate of $\epsilon_i$ by 
$\bar \epsilon ^i \equiv i (\epsilon_i)^\dagger \gamma^0$, 
the charge conjugation of $\epsilon _i$ is denoted as 
\begin{align}
\epsilon^i \equiv (\epsilon_i)^c =C (\bar \epsilon^i)^T =-i  \gamma^0 C(\epsilon_i)^* \,, 
\end{align}
where $C$ is the charge conjugation matrix satisfying $C^{-1}\gamma^\mu C=-\gamma_\mu ^T$. 
In this paper we adopt the representation such that the charge conjugation matrix is 
given by $C=-i \gamma^0$.  This enables us to raise and lower the indices $i,j,...$
simply by the complex conjugation and the gamma matrices are all real, hence
$\gamma_\mu ^T=\gamma^0 \gamma_\mu \gamma^0 $. 
It then follows that the spinors $\epsilon ^i$ with upper index 
have a positive chirality $\gamma_5 \epsilon^i =\epsilon^i$ and 
the Dirac conjugate of $\epsilon^i$ is given by 
$\bar \epsilon _i =- i (\epsilon^i)^\dagger \gamma^0$. 
Accordingly, the bilinears constructed out of the spinor satisfy 
\begin{align}
\bar \epsilon_i \epsilon_j =-\bar \epsilon_j \epsilon_i =(\bar \epsilon^i \epsilon^j)^* \,, \qquad 
i\bar \epsilon^i \gamma^\mu \epsilon_j=-i\bar \epsilon_j \gamma^\mu \epsilon^i \,, \qquad 
i \bar \epsilon_i\gamma_{\mu\nu} \epsilon_j =i \bar \epsilon_j\gamma_{\mu\nu} \epsilon_i \,.
\label{bilinear_prop}
\end{align}

Following the argument given in~\cite{NS}, 
 we define the supercovariant derivative operator as follows
 %---------------------- Supercovariant derivative   --------------------------%
\begin{align}
\hat \nabla_\mu \epsilon_i \equiv \nabla_\mu \epsilon_i +\mas A_{\mu i}{}^j \epsilon_j +
\mas B_{\mu ij}\epsilon^j \,, 
\end{align}
where $\nabla_\mu$ is an ordinary Lorentz-covariant derivative acting on a spinor.  
The $N\times N$ numbers of $4\times 4$ matrix-valued 
vector fields $\mas A_{\mu i}{}^j $  and $\mas B_{\mu ij}$ 
represent the deviation from the Levi-Civit\`a connection. 
These connections obey different commutation relations with 
the chirality matrix 
\begin{align}
[\mas A_{\mu i}{}^j , \gamma_5]=0 \,, \qquad \{
\mas B_{\mu ij}, \gamma_5\}=0 \,. 
\label{AB_comu}
\end{align} 
The Dirac conjugate of the supercovariant derivative is given by 
%---------------------- Supercovariant derivative: conjugate   --------------------------%
\begin{align}
 \overline{\hat \nabla_\mu \epsilon^i} = \overline{\nabla_\mu \epsilon^i}
-\bar \epsilon^j \gamma^0(\mas A_\mu{}^i{}_j)^T\gamma^0 +\bar \epsilon_j \gamma^0(\mas B_{\mu}{}^{ij})^T \gamma^0 \,. 
\end{align}
Here  the transpose operator $T$ is understood as acting on the space spanned by $4\times 4$ gamma matrices, 
whereas the raising and lowering the indices $i, j, ...$ are done by complex conjugation. 
We wish to put some constraints on the connections $\mas A_{\mu i}{}^j$ and 
$\mas B_{\mu ij}$ by requiring the positivity of energy.

Using the supercovariant derivative defined above, 
let us introduce the anti-symmetric Nester tensor~\cite{Nester}
%---------------------  Nester 2form   -------------------------%
\begin{align}
 N^{\mu\nu }=-i\left(\bar\epsilon ^{i}\gamma^{\mu\nu\rho }\hat \nabla_\rho \epsilon _i
-\overline{\hat \nabla_\rho \epsilon^i }\gamma^{\mu\nu\rho }\epsilon_i \right) \,, 
\label{Nester}
\end{align}
which reduces to the one in~\cite{Nester,NS} for $N=2$. 
The strategy employed by  Witten and Nester for the mass positivity is two-folds. 
Let us suppose that the asymptotically flat/AdS spacetime is foliated by some spacelike slice $\Sigma$. 
If $\Sigma$ is an orientable 3-surface, it turns out that the spacetime admits a  spin structure.  
This allows us to specify the appropriate fall-off rate of 
the metric, fluxes  and spinors on the spacelike surface $\Sigma$ in such a way that 
the following energy function is finite  and conserved 
\begin{align}
E_{\rm WN} = \frac 12 \int_{\partial \Sigma}  N_{\mu\nu}  \D S^{\mu\nu} \,, 
\end{align}
where $\partial \Sigma$ is the two-dimensional boundary of $\Sigma$ at infinity.  
In the asymptotically flat case, the Witten-Nester energy is related to the 
ADM momentum $P^\mu$~\cite{Arnowitt:1962hi} as $E_{\rm WN}=- V_{\infty}^\mu P_\mu$, where 
$V_\infty^\mu=i \bar \epsilon^i_\infty\gamma^\mu \epsilon_{\infty i}$
corresponds to the generator of the asymptotic translational symmetry and $\epsilon_{\infty i}$
are the asymptotic value of the spinors. 
The next step is to convert the surface integral at infinity--using the Stokes theorem--to the volume integral 
over $\Sigma$, 
\begin{align}
E_{\rm WN} =  \int_{\partial \Sigma}  \nabla_\nu N^{\mu\nu}  \D \Sigma_\mu \,,  
\end{align}
where $\D \Sigma_\mu$ is a past-directed volume element of $\Sigma$. 
If we can show $\nabla_a N^{0a}\ge 0$, where $0, a$ means the frame component, 
the Witten-Nester energy  turns out to be 
positive semi-definite $E_{\rm WN} \ge 0$. This leads to an inequality 
involving  globally conserved quantities such as mass, angular momentum, 
electromagnetic charges and so on.

Since the explicit form of $E_{\rm WN}$ is sensitive both to the asymptotic 
spacetime structures and to the field contents of the theory,  
we tentatively suppose that we can prescribe the boundary condition so that the 
Witten-Nester energy converges.  Hence our primary concern at the moment is 
the positivity of $\nabla_a N^{0a}$, or a lack thereof.  
After some computations,  the divergence of the Nester tensor 
can be brought into the following form, 
%--------------------------  2nd derivatives   -------------------------------%
\begin{align}
 \nabla_\nu N^{\mu\nu }=& 2i \overline{\hat \nabla_\rho \epsilon ^i}\gamma^{\mu\nu\rho }\hat
 \nabla_\nu \epsilon_i -G^\mu{}_\nu (i\bar \epsilon^i \gamma^\nu
\epsilon_i )
-\frac{i}{2}\bar \epsilon^i \left[ \gamma^{\mu\nu\rho }\mas
 F_{\nu\rho i}{}^j
+\gamma^0 (\mas F_{\nu\rho }{}^j{}_i)^T \gamma^0 
\gamma^{\mu\nu\rho }\right]\epsilon_j 
\nonumber \\
&-\frac{i}{2} \left[\bar \epsilon^i \gamma^{\mu\nu\rho }\mas
 H_{\nu\rho ij}\epsilon^j -\bar \epsilon_i \gamma^0 (\mas H_{\nu\rho
 }{}^{ji})^T \gamma^0 \gamma^{\mu\nu\rho }\epsilon_j \right]
\nonumber \\ & 
-i\bar \epsilon^i \left[\gamma^{\mu\nu\rho }\mas B_{\nu ik}\mas B_\rho
 {}^{kj}+\gamma^0 (\mas B_\nu {}^{jk}\mas B_{\rho ki} )^T \gamma^0
 \gamma^{\mu\nu\rho }\right] \epsilon_j 
\nonumber \\ &
+i\bar \epsilon^i [\gamma^{\mu\nu\rho }\mas A_{\nu i}{}^j
-\gamma^0 (\mas A_{\nu }{}^j{}_i)^T \gamma^0 \gamma^{\mu\nu\rho }
]\hat \nabla_\rho \epsilon_j 
-i\overline{\hat \nabla_\rho \epsilon^i }
[\gamma^{\mu\nu\rho }\mas A_{\nu i}{}^j
-\gamma^0 (\mas A_{\nu }{}^j{}_i)^T \gamma^0 \gamma^{\mu\nu\rho }
]\epsilon_j \nonumber \\ & 
-i\bar \epsilon_i[\gamma^{\mu\nu\rho }\mas B_{\nu}{}^{ij}-\gamma^0
 (\mas B_\nu {}^{ji})^T \gamma^0 \gamma^{\mu\nu\rho }]\hat \nabla_\rho
 \epsilon_j 
-i\overline{\hat \nabla_\rho \epsilon^i } \left[\gamma^{\mu\nu\rho }
\mas B_{\nu ij}-\gamma^0 (\mas B_{\nu ji})^T \gamma^0 \gamma^{\mu\nu\rho
 }\right]\epsilon^j \,, 
 \label{divN}
\end{align}
where 
we have defined the two kinds of curvatures 
%--------------------------  Curvatures -------------------------------%
\begin{subequations}
\begin{align}
 \mas F_{\mu\nu i}{}^j &= 2( \nabla_{[\mu }\mas A_{\nu ]i}{}^j 
+\mas A_{[\mu i}{}^k \mas A_{\nu ] k}{}^j) \,, \\
\mas H_{\mu\nu ij}&=2(\nabla_{[\mu}\mas B_{\nu] ij}+\mas A_{[\mu
 i}{}^k \mas B_{\nu]kj}+\mas B_{[\mu ik}\mas A_{\nu]}{}^k{}_j) \,. 
\end{align}
\end{subequations}
Readers should observe the following relation in deriving eqn.~(\ref{divN}), 
\begin{align}
i\bar \epsilon^i \gamma^{\mu\nu\rho} \mas B_{\nu ij}\hat \nabla_\rho \epsilon ^j=
(-i\bar \epsilon_i \gamma^{\mu\nu\rho} \mas B_{\nu}{}^{ij}\hat \nabla_\rho \epsilon _j)^\dagger 
= i \overline{\hat \nabla_\rho \epsilon^i} \gamma^0 (\mas B_{\nu ji})^T \gamma^0 \gamma^{\mu\nu\rho} \epsilon^j \,. 
\end{align}

We assume that the spinors $\epsilon_i$ satisfy the 
Dirac-Witten condition on $\Sigma$~\cite{Witten}, 
\begin{align}
\label{DW}
\gamma^I \hat \nabla_I \epsilon_i =0 \,, \qquad I=1,2,3\,. 
\end{align}
If there exist spinors satisfying this differential equation and  giving a 
finite Witten-Nester energy, 
the first term of the right side of (\ref{divN}) gives the nonnegative contribution 
to the volume integral due to 
$\overline{\hat\nabla_\rho\epsilon ^i}\gamma^{0\nu\rho }\hat\nabla_\nu \epsilon_i
=g^{IJ} (\hat \nabla_I\epsilon_i)^\dagger(\hat\nabla_J \epsilon_i)\ge 0$. 
According to our convention, the vector field $V^\mu \equiv i \bar \epsilon^i\gamma^\mu \epsilon_i $
is future-directed and nonspacelike  because of $V^0=\epsilon_i^{\dagger} \epsilon_i>0$. 
It follows that  the term $-G^\mu {}_\nu V^\nu$ turns out to have a positive contribution to the 
Witten-Nester energy, provided Einstein's equations hold and matter fields satisfy the 
suitable energy conditions. On the other hand, 
the last four terms proportional to $\hat \nabla_\rho \epsilon_i$ in eqn~(\ref{divN})
do not to have a definite sign. Hence we demand 
as a minimal requirement for the positivity of mass that the gauge 
connections should be subjected to the subsequent conditions, 
 %---------------------  Hermitian   -------------------------%
\begin{subequations}
\label{positivity}
\begin{align}
\gamma^0 (\mas A_\rho{}^j{}_i)^T \gamma^0 \gamma^{\mu\nu\rho }&=
 \gamma^{\mu\nu\rho }\mas A_{\rho i}{}^j \,, \label{positivity_A}\\ 
\gamma^0(\mas B_\rho {}_{ji})^T\gamma^0 \gamma^{\mu\nu\rho }&= \gamma^{\mu\nu\rho}\mas B_\rho{}_{ij}\,. \label{positivity_B}
\end{align}
\end{subequations}
We shall refer to these conditions as ``positivity conditions.''
Although we have not shown that these conditions are necessary, 
this requirement seems persuasive since all the theories which have been 
shown to admit the mass positivity in refs.~\cite{Gibbons:1993xt,Gibbons:1982fy,Gibbons:1982jg,Gibbons:1983aq,Boucher:1984yx,Townsend:1984iu,Gibbons:1994vm,Kostelecky:1995ei,Shiromizu:1999xm,Rogatko:2002qe}
indeed satisfy this property. 
Under the positivity conditions, the divergence of the Nester tensor takes 
a remarkably simple form 
%---------------------  divN   -------------------------%
\begin{align}
 \nabla_\nu N^{\mu\nu }=& 2i \hat \nabla_\rho \epsilon ^i \gamma^{\mu\nu\rho }\hat \nabla_\nu \epsilon_i 
-{G^\mu }_\nu V^\nu 
+S^\mu \,, 
\label{divNsol}
\end{align}
where the current $S^\mu =S_{(1)}^\mu+S_{(2)}^\mu +S_{(3)}^\mu$
is built out of  three different contributions, 
%------------------  Curvature 2form  -------------------%
\begin{subequations}
\begin{align}
S^\mu_{(1)}\equiv &-i\bar \epsilon^i \gamma^{\mu\nu \rho }\mas F_{\nu\rho
 i}{}^j\epsilon _j \,, \\ 
S^\mu_{(2)} \equiv  &
-\frac{i}{2} \left(\bar \epsilon^i \gamma^{\mu\nu\rho }\mas
 H_{\nu\rho ij}\epsilon^j -\bar \epsilon_i
\gamma^{\mu\nu\rho }\mas H_{\nu\rho }{}^{ij}
\epsilon_j \right)\,, \\
S_{(3)}^\mu\equiv &-2i \bar \epsilon^i \gamma^{\mu\nu\rho }
\mas B_{\nu ik}\mas B_{\rho }{}^{kj} \epsilon _j \,. 
\end{align}
\end{subequations}
Hence if we can show that the zero-th component of the current 
\begin{align}
\label{current}
  J^\mu \equiv -G^\mu{}_\nu V^\nu+S^\mu
\end{align}
 is nonnegative $J^0 \ge 0$ modulo the field equations, 
we can conclude the positivity of the Witten-Nester energy. 
Due to the simplicity of the formula (\ref{divNsol}), 
our approach can circumvent complications 
encountered in the model-dependent analysis.  

A possible way to find the gauge connections satisfying 
(\ref{positivity}) is to expand them in terms of the Clifford basis. 
We give the classification of the connections  in appendix. 
It turns out that the possible connections take the same form 
as those in extended supergravity if we impose an additional 
condition that $V^\mu =i \bar \epsilon^i \gamma^\mu \epsilon_i$
is a Killing field when $\hat \nabla_\mu \epsilon_i=0$ is satisfied. 
Note that this does not immediately imply that the Witten-Nester 
energy is positive and finite, since these conditions are not sufficient to
prove $J^0\ge 0$, and the finiteness of the surface integral is sensitive 
to the boundary conditions for the metric, gauge fields and scalars.

\section{Explicit examples}
\label{ex}

Exploiting the formulation developed in the previous section, 
we shall now demonstrate the positivity of the Witten-Nester energy
for various theories.  
The models we shall discuss are all motivated by extended supergravities. 
The following analysis illustrates that 
$\mas A_{\mu i}{}^j$ and $\mas B_{\mu ij}$ 
correspond respectively to the connection of the spinor bundle and 
to the contribution coming from the flux torsion. 
It turns out that the positivity conditions (\ref{positivity}) 
are indeed true for all models inspired by extended supergravities.

\subsection{$N=2$ minimal gauged supergravity}
\label{secN2}

Let us begin with the positivity of Witten-Nester energy in $N=2$ minimal gauged supergravity, 
i.e., the Einstein-Maxwell theory with a negative cosmological constant
\begin{align}
L=R -F_{\mu\nu}F^{\mu\nu} -2 \Lambda \,, 
\end{align}
where $F=\D A$ and $\Lambda =-3\ell^{-2}<0$. 
The field equations of this system are given by\footnote{
We do not consider here the extra source terms 
terms $T_{\mu\nu}^{(\rm mat)}$ and $J^\mu +i \ti J^\mu$ 
to the right side of Einstein's and Maxwell's equations, respectively.  
The positive mass property continues to be valid provided that  $T_{\mu\nu}^{(\rm mat)}$
satisfies the dominant energy condition, and that $J^\mu$ and $\ti J^\mu$ are future-pointing timelike vectors.
}

\begin{align}
G_{\mu\nu}+\Lambda g_{\mu\nu}=T^{(\rm em)}_{\mu\nu } \equiv 
2\left(
F_{\mu}{}^\rho F_{\nu\rho} -\frac 14g_{\mu\nu}F_{\rho\sigma}F^{\rho\sigma}
\right)
\,,  \qquad 
\nabla_\nu (F^{\mu\nu}+i  \star F^{\mu\nu})=0 \,.
\label{N2_EOM}
\end{align}
This subject was first discussed in \cite{Kostelecky:1995ei} by using a single Dirac spinor.
We demonstrate below that the argument in~\cite{Kostelecky:1995ei}
concerning the surface integral should be refined. 

The connections in the supercovariant derivative are given by 
\begin{align}
\mas A_{\mu i}{}^j = \frac i\ell (\sigma^3)_i{}^j   A_\mu  \,, \qquad 
\mas B_{\mu ij} =\frac 14 \epsilon_{ij}F_{\nu\rho}\gamma^{\nu\rho}\gamma_\mu
+\frac{i}{2\ell} (\sigma^3)_{ij}\gamma_\mu \,, 
\label{N2_AB}
\end{align}
where $(\sigma^I)_i{}^j$ is a standard Pauli matrix, whose index is  lowered 
by the alternate tensor $\epsilon_{ij}$  with $\epsilon_{12}=-\epsilon_{21}=1$ as 
$(\sigma^I)_{ij} \equiv \epsilon_{ki}(\sigma^I)_j{}^k$, viz
$(\sigma^3)_{ij}=(\sigma^1)_i{}^j$. Note that our convention leads to 
$(\sigma^I)^{ij}=[(\sigma^I)_{ij}]^*$ which differs from the  one  in \cite{Cacciatori:2008ek}.   
It is a simple exercise to verify that the connections (\ref{N2_AB}) obey the 
positivity conditions (\ref{positivity}) and (\ref{app_sol_AB}). Hence we get 
\begin{align}
S^\mu_{(1)}&= -\frac 2{\ell} \star F^{\mu\nu}(\sigma^3)_i{}^j 
(i \bar \epsilon^i \gamma_\nu \epsilon_j)\,, \nonumber \\
S^\mu_{(2)}&=(T^{({\rm em})\mu}{}_\nu -\Lambda \delta^\mu{}_\nu)V^\nu 
-S_{(1)}^\mu \,, 
\\
S^\mu_{(3)}&=\epsilon_{ij}(\nabla_\nu F^{\mu\nu}+i \nabla_\nu\star F^{\mu\nu})i\bar \epsilon^i\epsilon^j+{\rm c.c} 
\,, \nonumber 
\end{align}
thereby the current $J^\mu=-G^\mu{}_\nu V^\nu+S^\mu$ vanishes 
when the equations of motion~(\ref{N2_EOM}) are satisfied. 
This probes that the Witten-Nester energy is indeed positive semi-definite. 
It is worth commenting that the negativity of the cosmological constant is essential. 
An attempt to give a positive cosmological constant 
does not work, since the positivity conditions~(\ref{positivity}) fail to hold.\footnote{
Unlike in the Dirac spinor formulation in~\cite{NS}, 
the ``fake'' Killing  spinor equations for $\Lambda=3H^2>0$ are not obtained by the 
simple Wick-rotation $\ell \to i H^{-1}$ of (\ref{N2_AB}), 
since it is incompatible with raising and lowering the ${\rm SU}(2)$ indices
via complex conjugation. 
In the $\Lambda>0$ case, we have to choose 
$\mas A_{\mu i}{}^j=HA_\mu \delta_i{}^j $ 
and 
$\mas B_{\mu ij}=\epsilon_{ij}(\frac 14F_{\nu\rho}\gamma^{\nu\rho}\gamma_\mu+\frac 12H\gamma_\mu)$
in order to produce the correct equations of motion~\cite{Meessen:2009ma}. 
The latter connection does not satisfy the positivity conditions, as expected. 
We thank D. Klemm for useful comments about this. 
}
This would convince us that the positivity conditions~(\ref{positivity}) are 
indeed related to the mass positivity. 

The surface integral can be expressed in terms of globally conserved quantities as follows. 
It is convenient here to exploit the Dirac spinor $\eta = \epsilon^1-i \epsilon_2$ 
to evaluate the surface integral. Let us assume 
that the spacetime asymptotes to the AdS at infinity following the notion of refs.~\cite{Abbott:1981ff,Henneaux:1985tv,Ashtekar:1999jx,Hollands:2005wt}. We require that the Dirac spinor $\eta$ tends to the 
Killing spinor $\zeta$ of AdS at infinity and obeys
\begin{align}
\hat \nabla_\mu \eta =O(1/r^2) \,, \qquad 
\textrm{as $r\to \infty$}\,. 
\label{N2_BC}
\end{align}
The expression of Witten-Nester energy was derived in~\cite{Kostelecky:1995ei} and reads
\begin{align}
E_{\rm WN} =\bar \zeta J_{AB} \sigma^{AB} \zeta
- \bar \zeta(Q_e -i \gamma_5 Q_m ) \zeta\,, 
\label{N2_EWN}
\end{align}
where 
$\sigma^{AB}$ is the generator of ${\rm SO}(3,2)$ in the spinor representation and 
$J_{AB}$ is the ${\rm SO}(3,2)$  momentum ($A,B,=0,...,4$).  $Q_e$ and $Q_m$ denote  the 
electric and magnetic charges
defined by 
\begin{align}
\label{charges}
Q_e =\int_{\partial \Sigma} \star F, \qquad 
Q_m=\int _{\partial \Sigma} F\,.
\end{align}
Kostelecky and Perry~\cite{Kostelecky:1995ei} then 
concluded that the BPS bound should be  given by 
$M\ge \ell^{-1}|J|+\sqrt{Q_e^2+Q_m^2}$, where 
$M=J_{04}$ and $J=\ell J_{12}$ represent 
the mass and angular momentum~\cite{Abbott:1981ff,Henneaux:1985tv,Ashtekar:1999jx,Hollands:2005wt}. 
However, it has been pointed out in 
refs.~\cite{Romans:1991nq,Caldarelli:1998hg} that the magnetically charged 
Reissner-Norsdtr\"om-AdS solution
cannot be supersymmetric.

This apparent contradiction can be resolved in the following manner. 
The Killing spinor $\zeta$ in AdS satisfies 
$[\nabla_\mu+(1/2\ell)\gamma_\mu ] \zeta=0$ and 
is given by~\cite{Henneaux:1985tv}, 
\begin{align}
\label{AdS_KS}
\zeta =\Bigl(\cosh \frac{\rho}2 +\sinh \frac{\rho}2\gamma^1\Bigr)
\Bigl(\cos \frac{t}{2\ell} +\sin \frac{t}{2\ell} \gamma^0\Bigr)
\Bigl(\cos \frac{\theta}2 +\sin \frac{\theta}2\gamma^{12}\Bigr)
\Bigl(\cos \frac{\phi}2 +\sin \frac{\phi}2\gamma^{23}\Bigr)
\zeta_0\,, 
\end{align}
where $\zeta_0$ is a constant Dirac spinor. Here we have employed the 
global coordinates, 
\begin{align}
\D s^2 = -\cosh\rho^2 \D t^2 +\ell^2 [\D \rho^2 +\sinh^2\rho (\D \theta^2+\sin^2\theta \D \phi^2)] \,. 
\end{align}
The standard radial coordinate is given by $r=\ell \sinh\rho$. 
Given the explicit form of the Killing spinor~(\ref{AdS_KS}), 
one can compute the spinor bilinears appearing in the electric and magnetic charges
of~(\ref{N2_EWN}) as
\begin{subequations}
\begin{align}
\bar \zeta \zeta &=c_0\,, \\
  i\bar \zeta \gamma_5 \zeta &=
[c_1 \cos(t/\ell) +c_2 \sin(t/\ell)]\cosh\rho +
[c_3 \cos\theta +\sin\theta (c_4\cos\phi+c_5\sin\phi)]\sinh \rho \,, 
\end{align}
\end{subequations} 
where $c_{0-5}$ are real constants built out of $\zeta_0$. 
For the choice $c_{1-5}=0$, one can verify that 
$U_\mu\equiv i\bar\zeta\gamma_\mu\gamma_5\zeta=-\ell\nabla_\mu(i\bar\zeta\gamma_5\zeta)$ vanishes, 
which in turn implies that $V_\mu =i \bar \zeta \gamma_\mu \zeta$ is null due to the Fierz identity. 
This boundary condition is not the case that we are interested in, 
so at least one of $c_{1-5}$ is nonvanishing.  
It then follows that the bilinear $i\bar \zeta \gamma_5 \zeta$ appearing in the magnetic 
charge of~(\ref{N2_EWN}) diverges at infinity ($\rho \to \infty$), rendering the 
Witten-Nester energy~(\ref{N2_EWN}) ill-defined. 
Moreover, the presence of magnetic charge implies that the gauge potential $A_\mu$ 
cannot be globally defined, otherwise the integral of $Q_m$ in (\ref{charges})
vanishes. It is typically singular on the axis, leading to the source of delta-function.   
Since the gauge potential $A_\mu$ appears explicitly in the supercovariant derivative~(\ref{N2_AB}), 
the Dirac-Witten operator 
$\gamma^i \hat \nabla_i $ therefore cannot be straightforwardly invertible using Green's function. 
If one attempts to remove this distributional singularity,  the topological structure
of spacetime must change, resulting in a different 
BPS-bound~\cite{Hristov:2011ye,Hristov:2011qr}.
In the case of asymptotically globally AdS case in the Einstein-Maxwell-$\Lambda$ system, 
we therefore arrive at the inequality\footnote{
The appearance of angular momentum into the Witten-Nester energy can also be understood from the fact that 
in the framework of $N=2$ gauged supergravity, 
the bilinear vector field $V^\mu =i\bar \zeta\gamma^\mu \zeta$ in AdS 
is rotating by the constant angular velocity $\ell^{-1}$ with respect to the 
static observer at infinity~\cite{Klemm:2013eca}. 
} 
\begin{align}
M\ge \frac 1\ell |J| + Q_e \,. 
\end{align}
From the standpoint of ${\rm Osp}(4|2)$ superalgebra, 
the introduction of magnetic charge as central extension is forbidden 
since it fails to satisfy the Jacobi identity due to the breakdown of the ${\rm SO}(3,2)$ 
covariance~\cite{Dibitetto:2010sp}. 
Our explanation seems more convincing in the present context since the positive mass theorem 
does not assume the underlying supergravity theories in advance.

\subsection{K\"ahler target space}

Let us next discuss the case in which the set of the complex scalar fields
parameterizing the K\"ahler manifold  
is the source of Einstein's equations. Namely we shall 
concentrate on the theory
%---------------------  Action   -------------------------%
\begin{align}
L=R- 2G_{\alpha \bar \beta}g^{\mu\nu}\partial_\mu z^\alpha \partial_\nu \bar
 z^{\bar \beta} -V (z, \bar z )\,, \qquad
G_{\alpha \bar \beta }=\frac{\partial^2 K}{\partial z^\alpha \partial
 \bar z^{\bar\beta }}\,,
\end{align}
where $K(z, \bar z)$ is a real K\"ahler potential and $V(z,\bar z)$ is the 
potential to be determined by requiring the positivity of the Witten-Nester energy.
The indices $\alpha, \bar\beta$ run over any positive integers corresponding to the 
number of complex scalars.  
The Einstein equations following from this Lagrangian read
%---------------------  Einstei's eqs   -------------------------%
\begin{align}
G_{\mu\nu }=T_{\mu\nu } \,, \qquad 
T_{\mu\nu }=2 G_{\alpha \bar \beta }\left(\nabla_{(\mu }z^\alpha \nabla_{\nu )}\bar z^{\bar\beta} 
-\frac{1}{2}g_{\mu\nu }\nabla_\rho z^\alpha \nabla^\rho \bar z^{\bar
 \beta }\right) -\frac{1}{2}g_{\mu\nu }V \,.
\end{align}
We assume that $G_{\alpha \bar \beta}$ is a positive matrix, 
or equivalently the null energy condition, 
in such a way that there appear no ghosts.

We take $\mas A_{\mu i}{}^j$ and $\mas B_{\mu ij}$ 
satisfying the positivity condition (\ref{positivity})  and (\ref{app_sol_AB}) as
%---------------------  Kahler U(1)   -------------------------%
\begin{align}
(\mas A_\mu)_i{}^j = \frac 14  (K_{\alpha }\partial_\mu z^\alpha -K_{\bar \alpha }
\partial_\mu \bar z^{\bar \alpha })\delta_i{}^j \,, 
\qquad \mas B_{\mu ij}= \frac 12 e^{K/2}  W \gamma_\mu \delta_{ij} \,. 
\end{align} 
where $K_\alpha \equiv \partial K/\partial z^\alpha $. 
Here $W=W(z)$ is a holomorphic function of $z^\alpha $ and is referred to as a superpotential.
The superpotential is assumed to transform as $W\to W e^{-f}$ 
under the K\"ahler gauge transformation 
$K\to K+f+\bar f$.
The connection $\mas A_{\mu i}{}^j$ represents the ${\rm U}(1)^N$ connection.

We define the ``variation of dilatini'' as
%---------------------  dilatino   -------------------------%
\begin{align}
 \delta \lambda^\alpha{}_i=\delta_{ij} \gamma^\mu \partial_\mu z^\alpha \epsilon^j 
-e^{K/2} G^{\alpha \bar \beta} D_{\bar \beta}\bar W  \epsilon_i \,,  
\label{dilatini}
\end{align}
which has negative chirality $\gamma_5 \delta\lambda^\alpha{}_i=-\delta\lambda^\alpha{}_i$. 
$D_\alpha$ denotes the K\"ahler ${\rm U}(1)$ covariant derivative, which acts 
 on the superpotential  as  
\begin{align}
D_\alpha W =\partial_\alpha W +K_\alpha W \,. 
\end{align}

In this case, the on-shell current $J^\mu$ takes the manifestly nonnegative  form 
\begin{align}
J^\mu =-(G^\mu{}_\nu -T^\mu{}_\nu)V^\nu +
G_{\alpha \bar \beta} i\overline{\delta \lambda^{\beta i}}\gamma^\mu \delta 
\lambda ^\alpha {}_i \,, 
\end{align}
provided the potential is given by 
\begin{align}
\label{pot}
 V=2 e^{K}(G^{\alpha \bar \beta }  D_{\alpha }W D_{\bar \beta }\bar W
 -3|W|^2) \,, 
\end{align}
where $G^{\alpha \bar\beta}$ is the inverse of $G_{\alpha \bar \beta}$ and 
$\overline{\delta \lambda^{\beta i}}=i (\delta \lambda^\beta{}_i)^\dagger \gamma^0$. 
The existence of the superpotential obeying the desired transformation under the K\"ahler gauge transformation implies that the 1st Chern class of the line bundle coincides with the K\"ahler class, i.e., 
the manifold must be Hodge. 
The above discussion means that the volume integral is positive semi-definite
as far as the Hodge-K\"ahler target space is concerned, irrespective of 
the choice of superpotential.

The surface integral is finite if we impose eqn. (\ref{N2_BC}) for the spinors
and that the scalar fields fall off faster than $r^{-3/2}$.  
This boundary condition for the scalar implies that  
the mass eigenvalues should be above the Breitenlohner-Freedman bound
$m_{\rm BF}^2=-9/(4\ell^2)$~\cite{Breitenlohner:1982jf}, where 
$\ell $ is the curvature radius of the AdS vacua.\footnote{
It is important to note that 
if the mass of the scalar field is in the range $m^2_{\rm BF}\le m^2 \le m_{\rm BF}^2 +\ell^{-2}$, 
the slowly decaying solution is also normalizable, admitting any boundary conditions. 
In this case, 
it is unclear if the Witten-Nester energy coincides with 
other definitions of charges, e.g., the one introduced in~\cite{Breitenlohner:1982jf}.
See e.g, refs.~\cite{Hertog:2003xg,Hertog:2004dr} for the recent work addressing this problem. 
We content ourselves here by imposing the Dirichlet boundary conditions. 
}
The finiteness of the Witten-Nester energy is not guaranteed for the boundary condition employed 
in~\cite{Gibbons:1983aq}, in which case the negative mass initial data can be constructed 
along the line of~\cite{Hertog:2003xg}.

Despite the fact that the potential constructed from the K\"ahler potential and superpotential  is in general unbounded 
from below and above, 
the AdS vacua above the Breitenlohner-Freedman bound
are stabilized to allow the positive mass.
For example, the positivity of mass for the bosonic sector of the gravity multiplet in 
$N=4$ ${\rm SO}(4)$ gauged supergravity was discussed in  ref. \cite{Gibbons:1983aq}. In this case, 
the K\"ahler metric is given by the ${\rm SU}(1,1)/{\rm U}(1)$ coset and the 
superpotential takes a constant value~\cite{Gibbons:1983aq}, 
\begin{align}
K=- \ln (1-|\tau |^2) \,, \qquad 
W= \sqrt 2 g \,, 
\label{N4model}
\end{align}
where $\tau $ is an axidilaton and $g$ is an ${\rm SO}(4)$ gauge coupling constant. 
We can see that the potential is indeed unbounded from below and the origin is the unique vacuum
with the mass spectrum $m^2 =-2\ell^{-2}(\times 2)$. 
The analysis given in this section implies that the positivity of the mass 
holds in more general settings than the model considered in~\cite{Gibbons:1983aq}.  

It is worthwhile to comment that the stress energy tensor for the complex scalar 
field does not respect the dominant energy condition in general. 
The essential requirement that has played a crucial role here is  the null energy condition, 
viz., ${\rm eig}(G_{\alpha \bar \beta})\ge 0$~\cite{NS}.

It should be also noticed that the number $N$ of the Weyl spinors 
can be arbitrary.  One may be suspicious that this cannot be done since  
$N>8$ extended supersymmetric theory implies the necessity of introducing 
higher spin $s>2$ fields, which is a main obstacle  
to construct a local theory. 
However, the spinors only play a subsidiary role in the Witten-Nester formulation 
as the bilinear vector of the asymptotic symmetry. 
It therefore follows that the above argument actually has nothing to do with the full supergravity theories incorporating the fermion interactions even if it implies the underlying bosonic sector of supergravity theories. 
Hence our analysis continues to be valid also in the case involving $N>8$ spinors, and also in the 
(even) $D>11$  case, although its relevance to the physically interesting theories 
is less obvious.

\subsection{Einstein-Maxwell-dilaton theory}
\label{sec_EMD}

In this subsection, we consider the Einstein-Maxwell theory coupled to the dilaton field
with a potential, 
\begin{align}
L=R - 2(\nabla \phi)^2  -h(\phi) F_{\mu\nu} F^{\mu\nu} -2 V(\phi) \,, 
\end{align}
where $h(\phi)$ is the dilaton coupling function and $V(\phi)$ is the potential of the dilation, 
both of which are to be determined by requiring the positive mass. 
 The Einstein equations are given by 
 \begin{align}
G_{\mu\nu}=T_{\mu\nu} \,, \qquad 
T_{\mu\nu} =2 \left(\nabla_\mu \phi \nabla_\nu \phi -\frac 12 g_{\mu\nu
 }(\nabla \phi )^2 \right)-V g_{\mu\nu } +2h \left(F_{\mu \rho }F_\nu
 {}^\rho -\frac 14 g_{\mu\nu }F_{\rho \sigma }F^{\rho \sigma }\right) \,.
\end{align}
The dilaton and Maxwell equations read
\begin{align}
\nabla^2 \phi -\frac 12 V'(\phi) -\frac 14 h'(\phi) F_{\mu \nu
 }F^{\mu \nu }=0 \,, \qquad 
 \nabla_\nu [h(\phi)F^{\mu \nu }] =0 \,, \qquad \D F=0 \,.
 \label{EMD_eq}
\end{align}
Here the prime denotes the differentiation with respect to $\phi$. 

Setting $N=2$, we choose the gauge connections satisfying~(\ref{positivity}) and (\ref{app_sol_AB}) as follows
\begin{align}
\mas A_{\mu i}{}^j = -i g (\sigma^3)_i{}^j   A_\mu  \,, \qquad 
\mas B_{\mu ij} = \epsilon_{ij} K_1(\phi )F_{\nu\rho}\gamma^{\nu\rho}\gamma_\mu
+i W(\phi) (\sigma^3)_{ij}\gamma_\mu \,, 
\label{EMD_AB}
\end{align}
where $F=\D A$, $g$ is the gauge coupling constant and  
$K_1(\phi)$ and $W(\phi)$ are some real functions of $\phi$. 
We further define the variation of the spin $1/2$ fields as
\begin{align}
\delta \lambda_i = i \gamma^\mu \nabla_\mu \phi (\sigma^3)_{ ij}\epsilon^j 
+K_2(\phi) \epsilon_i 
+(\sigma^2)_i{}^j K_3(\phi)  F_{\mu\nu }\gamma^{\mu\nu} \epsilon_j   \,,  
\label{EMD_dilatini}
\end{align}
where $K_{2,3}(\phi)$ are again real functions. 
A straightforward computation shows that the on-shell current $J^\mu $
takes the nonnegative form, 
\begin{align}
J^\mu= -(G^\mu{}_\nu-T^\mu{}_\nu)V^\nu 
+i\overline{\delta \lambda^i}\gamma^\mu \delta \lambda _i
 \,, 
\end{align}
provided the Maxwell equations and the Bianchi identity $\D F=0$ hold, and 
if the following relations are satisfied 
\begin{subequations}
\label{EMD_sol}
\begin{align}
V=& K_2^2-12 W^2 \,, \qquad K_2=-2 W' \,, \label{EMD_sol1}\\ 
h =&4 (K_3^2+4 K_1^2) \,,  \qquad K_1 ^2 \propto h \,, \qquad K_3 =2 K_1 ' \,, \label{EMD_sol2}\\ 
0=& g+8WK_1 +2 K_2 K_3 \,. \label{EMD_sol3}
\end{align}
\end{subequations}
Equation  (\ref{EMD_sol1}) implies that the potential is expressed by the (real) superpotential as 
\begin{align}
V(\phi)=4 [W'(\phi)^2-3 W(\phi)^2] \,. 
\end{align}
The differential equation (\ref{EMD_sol2}) can be integrated to give 
\begin{align}
h=e^{-2 \alpha \phi } \,, \qquad 
K_1=\frac{1}{4 \sqrt{1+\alpha^2 }}e^{-\alpha \phi } \,, \qquad 
K_3=-\frac{\alpha }{2\sqrt{1+\alpha^2 }} e^{-\alpha \phi } \,,
\end{align}
where  $\alpha \in \mathbb R$ is the coupling constant of the dilaton. 
Finally, eqn.~(\ref{EMD_sol3}) is solved as 
\begin{align}
 W(\phi )=W_0 e^{-\phi/\alpha }-\frac{g }{2\sqrt{1+\alpha^2}} e^{\alpha\phi }\,,
\label{EMD_W}
\end{align}
where $W_0$ is the integration constant. 
Thus, in terms of the Dirac spinor $\eta =\epsilon^1-i \epsilon_2$, 
we have 
\begin{align}
 \hat \nabla_\mu \eta & =\left[\nabla_\mu +\frac{i}{4\sqrt{1+\alpha^2}}e^{-\alpha\phi }
F_{\nu\rho }\gamma^{\nu\rho }\gamma_\mu +W(\phi)\gamma_\mu +i g A_\mu \right] \eta \,,\label{EMD_KS1} \\
 \delta \lambda &=\left[\gamma^\mu \nabla_\mu \phi 
-2W'(\phi)-\frac{i \alpha }{2\sqrt{1+\alpha^2}}e^{-\alpha\phi }F_{\mu\nu }\gamma^{\mu\nu }\right]\eta \,. 
\label{EMD_KS2} 
\end{align}
When the potential vanishes, 
this recovers the result in~\cite{Gibbons:1993xt}. 
It is interesting that the superpotential and the dilaton coupling function $h(\phi)$ 
are completely determined by requiring the positivity within the class 
(\ref{EMD_AB}) and (\ref{EMD_dilatini}).  
If $W_0$ is tuned suitably, 
the above system with $\alpha=\pm \sqrt 3$ is obtained 
by the ${\rm U}(1)^4$ truncation of ${\rm SO}(8) $ maximal gauged supergravity~\cite{Duff:1999gh}. 

The Einstein-Maxwell-dilaton theory admitting the positive mass allows a 
free parameter $\alpha$, which characterizes how the 4-dimensional theory 
is derived from the higher-dimensional theory.  One may hope from the positive mass theorem above 
that this theory can be embedded into 
supergravity for an arbitrary value of $\alpha$. In order to see this, let us consider the 
BPS system described by~\cite{Nozawa:2010rf}
\begin{align}
\hat \nabla_\mu \eta =0 \,, \qquad \delta \lambda =0 \,. 
\label{EMD_BPS}
\end{align}
The first relation is first order differential equation, while the second is 
purely algebraic. Then it is not obvious for this BPS system to have a solution, thereby 
we have to check the integrability. 
Acting $\gamma^\nu \nabla_\nu$ to $\delta \lambda =0$ and using 
$\hat \nabla_\mu \eta =0$, 
we obtain 
\begin{align}
0= \Biggl[&
\nabla^2 \phi +4W' (3W-W'')+\frac{\alpha}{2}
e^{-2\alpha\phi }F_{\mu\nu }
F^{\mu\nu }+\frac{i e^{-\alpha\phi }}{\sqrt{1+\alpha^2 }}
\{\alpha (W-W'')+(\alpha^2-1)W'\}
\nonumber \\ & 
-\frac{i \alpha }{\sqrt{1+\alpha^2}} 
\{e^{-\alpha \phi }(\nabla_\mu \star F^{\mu\nu })\gamma_\nu \gamma_5 +
e^{\alpha\phi }\nabla_\mu (e^{-2\alpha\phi }F^{\mu\nu })\gamma_\nu \}
\nonumber \\&
 +\frac{i \gamma_5  \alpha (\alpha^2-3)}{2(1+\alpha^2 )}
e^{-2\alpha\phi }F_{\mu\nu }\star F^{\mu\nu }
\Biggr] \eta \,.
\label{EMD_int}
\end{align}
Assuming the dilaton field equation, the Maxwell equation and the Bianchi identity, 
the first and the second lines of (\ref{EMD_int}) drop out [note that the last term at the 
first line vanishes due to (\ref{EMD_W})]. However, the term at the third line 
remains nonvanishing unless $\alpha (\alpha^2-3)F\we F =0$. 
Hence it follows that the 1st-order system (\ref{EMD_BPS}) 
does not allow solutions in general except for $\alpha =0$ or $\alpha =\pm\sqrt 3$. The latter is obtained by the Kaluza-Klein reduction of 5-dimensional gravity if the potential is absent~\cite{Duff:1999gh}.   
This means that the general coupling case cannot be embedded into supergravity. 
Since the $F \we F \ne 0$ case corresponds to the dyonic metric, 
the purely electric/magnetic solution may admit Killing spinors as in the massless case~\cite{Nozawa:2010rf}.

\subsection{$N=8$ supergravity}

Finally, let us see the case of $N=8$ gauged supergravity. 
In ref. \cite{Gibbons:1983aq}, the positivity of the 
Witten-Nester energy was explored for the electric ${\rm SO}(8)$ gauging models
constructed by de Wit and Nicolai~\cite{de Wit:1982ig}. 
In recent years we have witnessed  a lot of progress in $N=8$ gauged supergravity. 
Of particular interest is the discovery of the one-parameter family of the deformation
of ${\rm SO}(8) $ gaugings~\cite{Dall'Agata:2012bb}
(see \cite{Kodama:2012hu} for the deformation of the ${\rm SL}(8)$-type gaugings). 
The deformed theory displays considerably rich physics compared to the undeformed one, 
since it admits new kinds of vacua~\cite{DI,Dall'Agata:2012sx} and 
new supersymmetry breaking patterns~\cite{DI,Catino:2013ppa}.  The deformation 
parameter might give rise to a new interpretation to M-theory embeddings and their 
field theory duals. Although the higher-dimensional origin of the noncompact gaugings is not identified yet, 
it has been extensively studied recently from the viewpoint of generalized geometry
(see e.g, \cite{Aldazabal:2013mya} and references therein). 
Hence it is intriguing to see whether the positivity of Witten-Nester energy 
depends on the deformation and the underlying gauging group. In particular, 
the noncompact gaugings might be associated to the ghost contribution, hence 
the positive mass property is quite nontrivial.  

The recent development of $N=8$ gauged supergravity  
is based on the embedding tensor formalism~\cite{de Wit:2007mt}, by which we can discuss 
in a duality covariant manner how to gauge a group by introducing additional 28 magnetic vector fields. 
The embedding tensor $\Theta_M{}^\alpha$ specifies how 
to choose the gauge group $G$ inside  $E_{7(7)}$, and 
defined by the relation $X_M=\Theta_M{}^\alpha t_\alpha$, 
where $t_\alpha$ and $X_M$ are the generators of $E_{7(7)}$ and $G$, respectively. 
Here $\alpha, \beta,...=1,...,133$ and $M,N,...=1,...,56$ are the adjoint and fundamental of $E_{7(7)}$.  
The consistent gaugings amount to requiring that the embedding tensor obeys 
linear and quadratic constraints~\cite{de Wit:2007mt}. The linear constraint implies that 
$\Theta_M{}^\alpha$ is sitting in the {\bf 912} representation of $E_{7(7)}$, whereas 
the quadratic constraint corresponds to the closure condition $\Omega^{MN}\Theta_M{}^\alpha \Theta_N{}^\beta=0$, 
where $\Omega^{MN}=i \sigma_2\otimes \mathbb I_{28}$ is the ${\rm Sp}(56, \mathbb R)$
invariant metric. 
Once the symplectic frame is chosen, the gauging can be done by the replacement 
$\partial_\mu \to D_\mu =\partial_\mu -g A_\mu{}^M\Theta_M{}^\alpha t_\alpha$, 
where $g$ is the gauge coupling constant and 
$A_\mu{}^M$ consists of electric and magnetic vectors 
$A_\mu{}^M=(A_\mu{}^\Lambda, A_{\mu\Lambda})$.

The Einstein's equations read~\cite{LeDiffon:2011wt} 
%----------------------  Einstein eqs   ----------------------------%
\begin{align}
 G_{\mu\nu }=T_{\mu\nu } \,, \qquad 
T_{\mu\nu }\equiv \frac{1}{6}\ma P_{(\mu}{}^{ijkl}\ma
 P_{\nu)ijkl}-\left(\frac{1}{12}
|\ma P_\rho |^2 +V \right)g_{\mu\nu } +\ma H^+{}_{(\mu }{}^\rho {}_{ij}
\ma H^-{}_{\nu)\rho }{}^{ij}\,, 
\label{N8_Ein}
\end{align}
where $\ma P_{\mu ijkl}$ is the self-dual vector field which corresponds to the 
kinetic term for scalars parameterizing the $E_{7(7)}/{\rm SU}(8)$ coset space, and 
given in terms of the mixed coset representative $\ma V_M{}^{\underline N}$
as $\ma P_{\mu ijkl}=i\Omega^{MN}\ma V_{Mij}\ma D_\mu \ma V_{Nkl}$, 
where $\ma D_\mu$ is the ${\rm SU}(8)$ covariant derivative~\cite{de Wit:2007mt}.  
The potential $V$ arises from the $O(g^2)$ corrections for the supersymmetry 
transformation and is constructed out of the $T$-tensor as~\cite{de Wit:2007mt}
\begin{align}
V=g^2 \left(
\frac 1{24}|A_{2i}{}^{jkl}|^2 -\frac 34|A_1{}^{ij}|^2 
\right) \,.  
\end{align}
Here $A_1$ and $A_2$ denote the 
${\bf 36}$ and ${\bf 420}$ irrep of the ${\rm SU}(8)$. 

The embedding tensor keeps the U-duality covariance at the price of 
introducing additional 28 magnetic vector fields $A_{\mu\Lambda}$, 
in addition to the usual electric vector fields $A_\mu{}^\Lambda$. 
This renders the usual field strength 
$\ma F_{\mu\nu}{}^M=2\partial_{[\mu}A_{\nu]}{}^M+g X_{[NP]}{}^MA_\mu{}^NA_\nu{}^P$
defined by the Ricci identity 
$[D_\mu,  D_\nu]=-g \ma F_{\mu\nu}{}^MX_M$
no longer covariant. A proposed prescription to overcome this is to 
introduce a tensorial auxiliary field $B_{\mu\nu \alpha}$ \cite{deWit:2005ub}, which
can be used to construct a covariant field strength 
$\ma H_{\mu\nu}{}^M=\ma F_{\mu\nu}{}^M+g Z^{M,\alpha}B_{\mu\nu\alpha}$,
where  $Z^{M,\alpha}\equiv \frac 12 \Omega^{MN}\Theta_N{}^\alpha$. 
Using the electric part of this field strength, it turns out that  
the following vector field strength transforms as a symplectic vector, 
\begin{align}
\ma G^+_{\mu\nu }{}^M
=
 \left(
\begin{array}{c}
 \ma H^+{}_{\mu\nu }{}^\Lambda \\
\ma N_{\Lambda\Sigma }\ma H^+{}_{\mu\nu }{}^\Sigma +2i \ma
 O^+{}_{\mu\nu \Lambda }
\end{array}
\right)
\,,
\end{align}
where 
$\ma N_{\Lambda \Sigma}=\ma N_{(\Lambda \Sigma)}$ is the kinetic term for the vector fields 
defined by $\ma V^{\Sigma ij} \ma N_{\Lambda\Sigma}=-\ma V_\Lambda{}^{ij}$, and 
$\ma O^+{}_{\mu\nu\Lambda}$ describes the fermion contribution~\cite{de Wit:2007mt}
which is taken to vanish in our computation, and ``$+$'' stands for 
the self-dual part, i.e., 
$\ma G^+_{\mu\nu }{}^M=\frac 12 (\ma G_{\mu\nu}{}^M-i \star \ma G_{\mu\nu}{}^M)$. 
The quantity $\ma H^+{}_{\mu\nu ij}$ appearing in Einstein's equation is dressed by 
a mixed coset representative as 
$\ma H^+{}_{\mu\nu ij}\equiv \ma V_{Mij}\ma G^+_{\mu\nu }{}^M$.
In terms of these ingredients,   
the equation for the vector fields is given by~\cite{LeDiffon:2011wt}
\begin{align}
E^\mu {}_{ij}\equiv  \ma D_\nu \ma H^{+ \mu \nu}{}_{ij}+\ma P_{\nu ijkl}\ma H^{-\mu\nu kl}
+\frac{g}{3} A_{2[i}{}^{nkl}\ma P^\mu{}_{j]nkl}=0 \,. 
\label{N8_Max}
\end{align}

We now turn to the discussion for the positive mass. 
As a connection $\mas A_{\mu i}{}^j $, we 
choose (half of) the ${\rm SU}(8)$ connection
$\mas A_{\mu i}{}^j=\frac 12 \ma Q_{\mu i}{}^j$ with
$\ma Q_\mu{}^i{}_j=-\ma Q_{\mu j}{}^i $ and 
$\ma Q_{\mu i}{}^i=0$.  
This connection obviously satisfies the positivity condition. 
Hence the current $S^\mu_{(1)}$ can be written in terms 
of the ${\rm SU}(8)$ curvature 
$F_{\mu\nu }(\ma Q)_i{}^j=2\partial_{[\mu }\ma Q_{\nu]}{}^i{}_j+\ma Q_{[\mu i}{}^k\ma Q_{\nu]k}{}^j$. 
Note that 
a physical degree of freedom is not encoded in this field,  since ${\rm SU}(8)$
is the maximal compact subgroup of $E_{7(7)}$. 
Using the Maurer-Cartan equations [see eqn. (3.5) of~\cite{LeDiffon:2011wt}], 
the ${\rm SU}(8)$ curvature is expressed by other fields and  
the current $S^\mu_{(1)}$
is given by 
\begin{align}
 S_{(1)}^\mu = 
-\frac{2}{3}\ma P_\nu {}^{jklm}\ma P_{\rho iklm}
(i\bar \epsilon^i \gamma^{\mu\nu\rho }\epsilon_j )
+g(\star \ma F^{\mu\nu M}) \ma Q_{Mi}{}^j 
\bar \epsilon^i \gamma_\nu \epsilon_ j \,,
\end{align}
where 
$\ma Q_{Mi}{}^{k}=\frac 23i \Omega^{NP}\ma V_{Nij}X_{MP}{}^Q\ma V_Q{}^{kj}$.

Let us take the connection $\mas B_{\mu ij}$ satisfying the positivity condition as 
%---------------------- B connection    --------------------------%
\begin{align}
\mas B_{\mu ij}=\frac{\sqrt 2}{8} \ma H^{+\rho\sigma}{}_{ij} \gamma_{\rho\sigma }\gamma_\mu
+\frac{g}{\sqrt 2} A_{1 ij}\gamma_\mu \,.  
\end{align}
The $T$-tensor variational identity [eqn. (D2) of~\cite{LeDiffon:2011wt}]  leads to 
$\ma D _\mu A_1{}^{ij}=-\frac 13 \ma P_\mu{}^{klm(i}A_2{}^{j)}{}_{klm}$, 
which yields 
\begin{align}
S_{(2)}^\mu 
=&
\sqrt 2 \left[
\ma D_\nu \ma H^{+\mu\nu }{}_{ij}
(i \bar \epsilon^i \epsilon^j ) -
\ma D_\nu \ma H^{-\mu\nu ij}(i \bar \epsilon_i
 \epsilon_j ) \right]
\nonumber \\
&+
\frac{{\sqrt 2}g}{3}(\ma P_{\nu klm(i} A_{2j)}{}^{klm}i \bar \epsilon^i \gamma^{\mu\nu }\epsilon^j 
-\ma P_\nu {}^{klm(i} A_2{}^{j)}{}_{klm}i \bar \epsilon_i
 \gamma^{\mu\nu }\epsilon_j )\,. 
 \label{N8_S2}
\end{align}
A simple computation shows that 
\begin{align}
 S_{(3)\mu} =&
-2 \ma H^+{}_{(\mu }{}^\rho {}_{ik}\ma H^-{}_{\nu)\rho }{}^{kj}
(i \bar \epsilon^i \gamma^\nu \epsilon_j )
+6 g^2 A_{1ik}A_1{}^{kj} (i\bar \epsilon^i \gamma_\mu \epsilon_j ) 
\nonumber \\
& 
+2g [A_{1ik}\ma H^{-}{}_{\mu \nu }{}^{kj}(i\bar \epsilon ^i \gamma^\nu \epsilon_j )-
\ma H^{+}{}_{\mu\nu }{}_{ik}A_1{}^{kj}(i\bar \epsilon ^i \gamma^\nu
 \epsilon_j)]\,.
 \label{N8_S3}
\end{align}

Finally we define the variation of dilatini as 
\begin{align}
\delta \chi_{ijk}&=-2\sqrt 2 \ma P_{\mu ijkl}\gamma^\mu \epsilon^l 
+\frac{3}{2} \gamma_{\mu\nu }\ma H^{+\mu\nu}{}_{ [ij}\epsilon_{k]}
-2 g A_{2}{}^l{}_{ijk}\epsilon_l \,.
\end{align}
The self-dual property of $\ma P_{\mu ijkl}$ implies 
$ \ma P_{(\nu}{}^{ijkl}\ma P_{\rho) ijkm}=\frac{1}{8}\ma P_{\nu ijkn}\ma P_{\rho}{}^{ijkn }\delta_m{}^l$,
hence after some calculations we find
\begin{align}
 i \overline{\delta \chi^{ijk}}\gamma^\mu \delta \chi_{ijk}
=& -18 (\ma H^{+\mu\rho }{}_{ij}\ma H^-{}_{\nu\rho }{}^{[ij}
+\ma H^{+}{}_{\nu\rho }{}_{ij}\ma H^{-\mu \rho [ij}
)
i \bar \epsilon^{k]}\gamma^\nu\epsilon_k
\nonumber \\
& -12
\sqrt 2[\ma H^{+\mu\nu }{}_{ij}\ma P_\nu{}^{ijkm}(i\bar \epsilon_m
 \epsilon_k )+\ma H^{-\mu\nu ij}\ma P_{\nu ijkm}(i\bar \epsilon^k
 \epsilon^m )]
\nonumber \\
& -12 g [\ma H^{+\mu }{}_{\nu ij} A_{2m}{}^{ijk}(i \bar \epsilon^m
 \gamma^\nu \epsilon_k )+\ma H^{-\mu }{}_\nu {}^{ij}A^m{}_{ijk}
(i \bar \epsilon^k \gamma^\nu \epsilon_m )]
\nonumber \\ & 
+8i \bar \epsilon^l \gamma^{\mu\nu\rho }\epsilon_m \ma P_{\nu ijkl}\ma P_\rho{}^{ijkm}
+4 g^2 A_{2l}{}^{ijk}A_2{}^m{}_{ijk}i \bar \epsilon^l \gamma^\mu 
\epsilon_m 
\nonumber \\ 
&- \left(\ma P^{\mu ijkl}\ma P_{\nu ijkl}+\ma P^\mu{}_{ijkl}\ma P_\nu {}^{ijkl}
-{\delta^\mu }_\nu |\ma P |^2 \right) i \bar \epsilon^m \gamma^\nu
 \epsilon_m 
\nonumber \\ & 
+4\sqrt 2 g  (i \bar \epsilon^l \epsilon^m \ma P^{\mu}{}_{ijkm}A_{2l}{}^{ijk} 
+i \bar \epsilon_l\epsilon_m \ma P^\mu {}^{ijkl}
A_2{}^m{}_{ijk})
\nonumber \\ &
+4\sqrt 2 g ( i\bar \epsilon^l\gamma^{\mu\nu } 
\epsilon^m \ma P_\nu {}_{ ijkm}A_{2l}{}^{ijk}- 
i \bar \epsilon_l\gamma^{\mu\nu }\epsilon_m \ma P_{\nu}{}^{ijkl}
A_2{}^m{}_{ijk})\,.
\label{dilatinisq}
\end{align}
Focusing on terms proportional to $i\bar \epsilon^i \gamma^\nu \epsilon_j$
in (\ref{N8_S3}) and (\ref{dilatinisq}), 
the following relation holds
\begin{align}
&\ma H^{+}{}_{\mu\nu kl}(A_{2i}{}^{jkl}+2 A_1{}^{j[k}\delta^{l]}{}_i)
+\ma H^-{}_{\mu\nu }{}^{kl}(A_2{}^j{}_{ikl}+2 A_{1i[k}\delta_{l]}{}^j)
\nonumber \\
&= -\frac{4}{3} (\ma H^{+}{}_{\mu\nu kl} T_i{}^{jkl}+\ma H^-{}_{\mu\nu
 }{}^{kl}
T^j{}_{ikl} ) =-i \Omega^{MN}\ma Q_{Mi}{}^j 
(\ma G^+{}_{\mu\nu }{}^P \ma V_{Pkl}\ma V_{N}{}^{kl}
+\ma G^-{}_{\mu\nu }{}^P \ma V_P {}^{kl}\ma V_{Nkl}) \nonumber \\ 
&=-\Omega^{MN}\Omega _{PN}\ma Q_{Mi}{}^j (\ma G^+{}_{\mu\nu }{}^P-\ma G^-{}_{\mu\nu
 }{}^P)
=i \ma Q_{Mi}{}^j \star \ma G_{\mu\nu }{}^M\,, 
\end{align}
where we have used 
$\ma V_M{}^{ij}\ma V_{Nij}-\ma V_{Mij}\ma V_N{}^{ij}=i\Omega_{MN}$, 
$\ma G^+{}_{\mu\nu }{}^P\ma V_P{}^{kl }=-\frac 12 \ma O^{+}{}_{\mu\nu}{}^{kl}$ and 
$\Omega ^{MN}\Omega _{PN}=\delta_P{}^M$. 
Due to the property $Z^{M ,\alpha }X_M=0$, 
we have 
$ \star (\ma F^M - \ma G^M)_{\mu\nu}\ma Q_{Mi}{}^j = \star (\ma H^M - \ma G^M)_{\mu\nu}\ma Q_{Mi}{}^j=0$, 
where the second equality follows from the equations of motion of $B_{\mu\nu\alpha}$.  
Combined with the fact that 
the $T$-tensor identity [eqn. (3.30) of~\cite{de Wit:2007mt}]  implies
\begin{align}
V\delta_l{}^m=g^2 \left(
 \frac{1}{3}A_{2l}{}^{ijk}A_{2}{}^m{}_{ijk}-6 A_{1li}A_1{}^{mi}\right) \,, 
\end{align}
it follows that 
the terms involving $i \bar \epsilon^i \gamma^\nu \epsilon_j$ are canceled out except for 
the stress energy tensor. We therefore arrive at
\begin{align}
S^\mu=T^\mu{}_\nu (i \bar \epsilon^i \gamma^\nu \epsilon_i)
+\frac{1}{12}
i \overline{\delta\chi^{ijk}}\gamma^\mu \delta \chi_{ijk} 
+\sqrt 2 [ (i \bar \epsilon^i \epsilon^j ) E_{ij}{}^\mu -(i\bar
 \epsilon_i \epsilon_j) E^{ij\mu }] \,.
 \label{N8_S}
\end{align}
This is the desired one which gives rise to the positive contribution to the
Witten-Nester energy when the bosonic equations of motion (\ref{N8_Max}) are satisfied, that is, 
the on-shell current $J^\mu$ becomes
\begin{align}
J^\mu=\frac{1}{12}i \overline{\delta\chi^{ijk}}\gamma^\mu \delta \chi_{ijk}  \,. \label{N8_J}
\end{align} 

In deriving eqn.~(\ref{N8_S}), we have used the 
Maurer-Cartan equations and the $T$-tensor identities. 
The Maurer-Cartan equation is derived based upon the closure relation 
$[X_M, X_N]=-X_{MN}{}^PX_P$, whereas the $T$-tensor identities 
are coming from the branching of {\bf 912} of $E_{7(7)}$ into
irrep of ${\rm SU}(8)$. This means that the positivity of Witten-Nester
energy holds as long as the linear and quadratic constraints on the
embedding tensor are satisfied (but the explicit solutions for these 
constraints are unnecessary). 
Namely, the positivity property continues to be valid for the 
consistent gaugings for any symplectic frames. 
This is a generalization of the result in~\cite{Gibbons:1983aq},
where the positivity has been shown for the ${\rm SO}(8)$ electric gaugings
of de Wit and Nicolai~\cite{de Wit:1982ig}.

\section{Summary}
\label{summary}

Inspired by the recent sparkling development of our understanding the extended supergravities, 
this article studied the positivity of mass in these theories. 
We presented a formulation for the positivity of Witten-Nester energy 
in terms of Weyl spinors. 
We found that the positivity conditions~(\ref{positivity}) should be satisfied as a 
minimal requirement for the positivity. These conditions are the direct generalization of 
the one proposed in our previous paper~\cite{NS}. 

We derived the universal formula (\ref{divNsol}) under the positivity conditions. 
Of particular use of this formula is its simplicity, allowing one to evaluate the 
mass positivity without lengthy computations as have been done in the literature. 
Although we have explored the ``positivity conditions'' for particular theories
inspired by supergravity, we have verified that
this is indeed true for all ungauged models in~\cite{Meessen:2010fh}. 
We gave a detailed proof for the classification of the connection in appendix. 
If we required that the bilinear vector field is a Killing vector for BPS states, 
it turned out that 
the possible connections take the same form as those appearing in 
extended supergravities (except for the unusual type of ``trombone gaugings'').  
There should presumably be a profound reason for this. 
We leave the deeper investigation for future study.

We revealed various new aspects that have been overlooked in the past studies
and provided a generalization of the mass positivity proof considered in the literature. 
We first revisited the minimal $N=2$ gauged supergravity, for which the 
contribution of the magnetic charge to the BPS-inequality was reconsidered. 
We argued the absence of magnetic charge without resorting the supersymmetry algebra. 
We expect that 
the similar argument can be carried out for the matter-coupled $N=2$ supergravity~\cite{Cacciatori:2008ek}.
As a generalization of the result in~\cite{Gibbons:1983aq}, 
it was shown that the positivity holds as far as the 
target space of the complex scalar is the Hodge-K\"ahler. 
This is a gratifying result from the viewpoint of scalar-multiplet in supergravity. 
In Einstein-Maxwell-dilaton theory, we showed that the dilaton coupling function and the 
superpotential are severely constrained due to the positive mass property. 
The supergravity embedding was explored by investigating the integrability condition for the 
dilation variation, allowing us to find that this is the case for the particular values of the 
coupling constant.  
We also extended the result of ref.~\cite{Gibbons:1983aq}
concerning the $N=8$ gauged supergravity by making use 
of the modern formulation based upon the embedding tensor. 
Recent development of the maximal gauged supergravity 
revealed that the deformed theories display interesting physics 
quite different from those predicted in undeformed theory. Despite that the 
positive mass property is obscure for deformed theories and 
for noncompact gaugings, 
we nevertheless demonstrated that the mass positivity is insensitive to the 
gauging and deformation parameter, as far as the 
linear and quadratic constraints on the embedding tensor 
are satisfied.

Recently, several gravitational theories have been considered motivated by 
dark energy. Most of these theories are phenomenological and suffer from 
various stability problems.  These theories may be constrained by requiring the 
positive mass, as discussed in~\cite{NS, Elder:2014fea}. For these purposes, 
the results of section~\ref{sec:PMT} and appendix would be of great help, 
since the possible connections are
highly restricted. Looking for modified gravitational theories admitting the positive mass and 
the (bosonic sector of) supergravity with noncanonical scalar fields are interesting future work. 
We hope to report the results in a separate paper.

\section*{Acknowledgement} 

MN wishes to thank  Dietmar Klemm and  Hideo Kodama for helpful discussions. 
MN  also appreciates the participants of the workshop ``Quantum Gravity, Black Holes and Strings'' 
for valuable comments and  the kind hospitality of Kavli Institute for Theoretical Physics China (KITPC).
TS thanks Akihiro Ishibashi for useful conversation. 
The work of MN is partially supported by JSPS research abroad and INFN. 
TS is supported by Grant-Aid for Scientific 
Research from Ministry of Education, Science, Sports and Culture of Japan (Nos.~21244033 and 25610055).

\appendix

\section{Classifying the positivity conditions}

In the body of text, we imposed the conditions (\ref{positivity}) 
on the connections in the supercovariant derivatives 
for the positivity of the Witten-Nester energy.  
Here we give a classification of the connection satisfying the positivity conditions (\ref{positivity}). 

Our strategy here is to expand the connections in terms of the 
Clifford basis 
$\{\mathbb I,\gamma_5, \gamma_\mu ,\gamma_\mu \gamma_5, \gamma_{\mu\nu }\}$. 
Taking into account the commutation relation~(\ref{AB_comu}), 
the connections $\mas A_{\mu i}{}^j$ and $\mas B_{\mu ij}$ can be 
expanded as\footnote{\label{footnote}
Expressions (\ref{app_A_ex}) and (\ref{app_B_ex}) are actually redundant, 
since $\mas A_{\mu i}{}^j$ ($\mas B_{\mu ij}$) acts on the spinors with 
negative (positive) chirality. Hence $a_{(2)}$  and $b_{(2)}$ can be absorbed respectively into 
$a_{(1)}$ and $b_{(1)}$, and  $a_{(3)}$ can be chosen to satisfy
$\frac 12\epsilon_{\nu\rho}{}^{\sigma\tau}a_{(3)\mu\sigma\tau}=ia_{(3)\mu\nu\rho}$. 
If this is done, however, the positivity conditions (\ref{positivity_A}) 
and (\ref{positivity_B})  must be projected by $1-\gamma_5$ and $1+\gamma_5$, 
respectively. In this case, one must take great care of the dual of the form fields. 
In order to circumvent this,  we leave the chiral matrix in (\ref{app_A_ex}) and (\ref{app_B_ex}), for which 
the basis $\{\mathbb I,\gamma_5, \gamma_\mu ,\gamma_\mu \gamma_5, \gamma_{\mu\nu }\}$
is independent. 
The redundancy can be removed by taking $a_{(2)}\to -a_{(1)}$, 
$b_{(2)} \to b_{(1)}$, 
$\frac 12 \epsilon_{\nu\rho}{}^{\sigma\tau}\ti a_{(3)\mu\sigma\tau}=i \ti a_{(3)\mu\nu\rho}$
and 
$a_{(3)}{}^\rho{}_{\rho \mu}=-\frac i2\epsilon_{\nu\rho\sigma\mu}a_{(3)}{}^{[\nu\rho\sigma]}$
[see (\ref{app_a3_dec}) for definition] at the final expression. 
Because of this, 
the imaginary self-dual property of $b_{(1)[\mu\nu]}$ follows from (\ref{app_bsol}).  
}
\begin{align}
    \mas A_{\mu i}{}^j=& a_{(1)\mu i}{}^j \mathbb I+ a_{(2)\mu i}{}^j  \gamma_5 +a_{(3)\mu\nu\rho i}{}^j \gamma^{\nu\rho} \,,   \label{app_A_ex}\\
\mas B_{\mu ij}  = & b_{(1)\mu\nu ij} \gamma^\nu +b_{(2)\mu\nu ij}\gamma^\nu \gamma_5 \,, 
\label{app_B_ex}  
\end{align}
where $a_{(1-3)}$ and $b_{(1-2)}$ are $N\times N$ matrix-valued tensorial fields with 
$a_{(3)\mu\nu\rho}=a_{(3)\mu[\nu\rho]}$. 

Let us begin with the case of $\mas B_{\mu ij}$. Substituting~(\ref{app_B_ex}) 
into (\ref{positivity}), expanding again by the Clifford basis and comparing the 
coefficients of the both sides of equation, 
we can get two set of relations 
\begin{align}
& b_{(1)[\mu \nu] ij}  =  -b_{(1)[\mu \nu] ji}\,, \qquad 
 b_{(2)[\mu \nu] ij} =- b_{(2)[\mu \nu] ji} \,,    \label{app_beq1}\\
 & b_{(1)\rho}{}^{[\mu }{}_{ij}\delta^\nu{}_{[\tau}\delta^{\rho]}{}_{\lambda]}
 +\frac i2 b_{(2)\rho}{}^{[\mu}{}_{ij}\epsilon^{\nu \rho]}{}_{\tau \lambda} 
 =  b_{(1)\rho}{}^{[\mu }{}_{ji}\delta^\nu{}_{[\tau}\delta^{\rho]}{}_{\lambda]}
 +\frac i2 b_{(2)\rho}{}^{[\mu}{}_{ji}\epsilon^{\nu \rho]}{}_{\tau \lambda} \,.
 \label{app_beq2}
\end{align} 
Contracting indices of (\ref{app_beq2}) and using (\ref{app_beq1}), 
we obtain 
\begin{align}
b_{(1)(\mu\nu) ij}=b_{(1) (\mu\nu)ji} \,, \qquad 
b_{(2)(\mu\nu) ij}=b_{(2) (\mu\nu)ji}  \,, \qquad 
b_{(1)[\mu\nu] ij}=- \frac i2 \epsilon _{\mu\nu\rho\sigma} b_{(2)}{}^{\rho \sigma}{}_{ij} \,.  
\label{app_bsol}
\end{align}
$b_{(1,2)(\mu\nu)}$ can be further decomposed into 
trace and trace-free parts as 
\begin{align}
b_{(I)(\mu\nu)ij}=\frac 14 g_{\mu\nu} b_{(I)\rho}{}^\rho {}_{ij} 
+\hat b_{(I)(\mu \nu) ij} \,, \qquad \hat b_{(I)\rho}{}^\rho {}_{ij} =0 \,, \qquad I=1,2\,. 
\end{align}

One can similarly obtain the relation for the coefficients $a_{(1-3)}$ as above. 
Suppressing the indices $i,j,..$, the 3-tensor $a_{(3)\mu \nu\rho}$ has 24 components. 
Hence it is decomposable into the irreducible parts 
${\bf 24}\to{\bf 4}+{\bf 16}+{\bf 4}$ as
\begin{align}
a_{(3)\mu \nu\rho} =a_{(3)[\mu\nu\rho]}+ \ti a_{(3)\mu\nu\rho} -\frac 23 a_{(3)}{}^\sigma{}_{\sigma [\nu}g_{\rho]\mu} 
\,, \label{app_a3_dec}
\end{align}
where 
$\ti a_{(3)\mu\nu \rho } \equiv 
\frac{2}{3}(a_{(3)\mu\nu\rho }-a_{(3)[\nu\rho ]\mu }+
a_{(3)\sigma }{}^\sigma {}_{[\nu }g_{\rho]\mu })$
satisfies 
\begin{align}
\ti a_{(3)\mu\nu\rho}=\ti  a_{(3)\mu[\nu\rho]} \,, \qquad 
\ti a_{(3)}{}^\sigma{}_{\sigma \mu}=\ti a_{(3)}{}^\sigma{}_{\mu\sigma}=0 \,, \qquad
\ti a_{(3)[\mu\nu\rho]}=0 \,. \label{app_tia3}
\end{align}
Noting that $\gamma_5$ is pure-imaginary and anti-symmetric in our convention,  
insertion of (\ref{app_A_ex}) into (\ref{positivity}) yields 
\begin{align}
\label{}
&a_{(3)[\mu\nu\rho] i}{}^j=-a_{(3)[\mu\nu\rho]}{}^j{}_i \,, \qquad 
\ti a_{(3)\mu\nu\rho i}{}^j=0 \,, \qquad 
a_{(3)}{}^{\rho}{}_{\rho \mu}{}_i{}^j = a_{(3)}{}^\rho{}_{\rho \mu }{}^j{}_i \,,  
\nonumber \\
&
a_{(1)\mu i}{}^j +a_{(1) \mu}{}^j{}_i =-\frac 43 a_{(3)}{}^\rho{}_{\rho \mu i}{}^j \,, \qquad  
a_{(2)\mu i}{}^j +a_{(2) \mu}{}^j{}_i =-\frac 23 i \epsilon_{\nu\rho \sigma \mu} a_{(3)}{}^{[\nu
\rho\sigma]}{}_i{}^j \,.\label{app_asol}
\end{align}

Equations~(\ref{app_beq1}), (\ref{app_bsol}) and (\ref{app_asol}) are 
exhaustive constraints arising from the positivity conditions~(\ref{positivity}) 
(see also the comments in footnote~\ref{footnote}). 
One can easily verify that all connections considered in the body of text satisfy 
these relations. 
Comparing with the model of Einstein-Maxwell-dilaton theory in section~\ref{sec_EMD}, 
one sees that $b_{(1)[\mu\nu][ij]}$ term correspond to the Maxwell field, 
$a_{(1)\mu i}{}^j$ is the gauge connection and 
$b_{(1)\rho}{}^\rho{}_{(ij)}$ denotes the 
superpotential contributions. 

Although the positivity conditions (\ref{positivity}) put some restrictions to 
the possible form of the connections, some unfamiliar terms 
($a_{(3)}{}^\rho{}_{\rho \mu}$ and $\hat b_{(1)\mu\nu}$) remain. 
Equation~(\ref{app_asol}) implies that $a_{(1)}$ fails to describe the connection 
contained in the subgroup of ${\rm U}(N)$ if $a_{(3)}{}^\rho{}_{\rho \mu}$ is nonvanishing.  
Also, there exist no supergravity models which contain 
$\hat b_{(1)(\mu \nu) ij}=\hat b_{(1)(\mu \nu) (ij)}$ (see e.g, \cite{Meessen:2010fh}
for ungauged models). 
Hence the positivity conditions leave some more freedom 
than extended supergravity models, 
although it is not clear yet such terms in fact produce
the {\it positive and finite} Witten-Nester energy. 

Nevertheless, 
we can fix these remaining terms as follows. 
 Let us consider the case in which $\hat \nabla_\mu \epsilon_i=0$
is satisfied, for which the spacetime is in ``BPS.''  
If the supergravity embedding is indeed possible,  
the bilinear vector $V^\mu =i\bar \epsilon^i \gamma^\mu \epsilon_i$ 
turns out to be a Killing field for the BPS metric~\cite{Maeda:2011sh}.\footnote{
The Einstein-Maxwell-dilaton theory does not have a supergravity origin for the 
general coupling as shown in section~\ref{sec_EMD}, 
yet this property continues to hold and the positivity condition is 
also met. In the Einstein-$\Lambda (>0)$ system for which the positivity condition is 
not satisfied, the bilinear vector field also	
fails to be a Killing vector. }  
Hence it might be reasonable to require that 
$V^\mu=i\bar \epsilon^i \gamma^\mu \epsilon_i$ satisfies the Killing equation when 
$\hat \nabla_\mu \epsilon_i =0$ is satisfied. This gives
\begin{align}
0=\nabla_{(\mu}V_{\nu)} =i \bar \epsilon ^i
[\gamma^0 (\mas A_{(\mu}{}^j{}_i)^T \gamma^0 \gamma_{\nu)}
-\gamma_{(\nu} \mas A_{\mu) i}{}^j] \epsilon_j 
+i \bar \epsilon_i \gamma_{(\mu}\mas B_{\nu)}{}^{ij}\epsilon_j 
-i \bar \epsilon^i \gamma_{(\mu }\mas B_{\nu ) ij} \epsilon^j \,,
\label{app_Killing}
\end{align}
where we have used 
$i\bar \epsilon_i\gamma^0(\mas B_{(\mu}{}^{ji})^T\gamma^0 \gamma_{\nu)}\epsilon_j
=-i \bar \epsilon_i \gamma_{(\nu}\mas B_{\mu)}{}^{ij}\epsilon_j $. 
It follows that $\mas A_{\mu i}{}^j$ obeys 
\begin{align}
\gamma^0 (\mas A_{(\mu}{}^j{}_i)^T \gamma^0 \gamma_{\nu)}=\gamma_{(\nu} \mas A_{\mu) i}{}^j \,.
\end{align}
Substituting (\ref{app_A_ex}) into the above equation and using~(\ref{app_asol}), 
we have further constraints 
\begin{align}
a_{(3)\rho}{}^\rho{}_{\mu i}{}^j=0 \,, \qquad 
a_{(3)[\mu\nu \rho]i}{}^j=0 \,. 
\end{align}
This implies that   
$a_{(1) i}{}^j=-a_{(1)}{}^j{}_i$, viz, $a_{(1)}$ is anti-hermitian 
and therefore describes the connection contained in ${\rm U}(N)$.

For the connection $\mas B_{\mu ij}$, eqn. (\ref{app_Killing}) does not 
imply  $\gamma_{(\mu }\mas B_{\nu ) ij}=0$ since 
the property (\ref{bilinear_prop}) must be taken into account. With this 
remark in mind, the condition $\bar \epsilon^i\gamma_{(\mu }\mas B_{\nu ) ij}\epsilon^j=0$
yields 
\begin{align}
\hat b_{(I)(\mu\nu) ij}=0 \,, \qquad I=1,2\,.  
\end{align}

After the replacement $a_{(2)} \to -a_{(1)}, b_{(2)} \to b_{(1)}$
with chiral projections, 
we finally arrive at 
\begin{align}
\label{app_sol_AB}
\mas A_{\mu i}{}^j =A_{\mu i}{}^j \mathbb I \,, \qquad 
\mas B_{\mu ij}= W_{(ij)} \gamma_\mu 
+F_{\mu\nu [ij]} \gamma^\nu \,, 
\end{align} 
where $A_\mu$ is anti-hermitian, 
$W_{(ij)}$ is an $N\times N$ symmetric matrix 
and $F_{\mu \nu}=F_{[\mu\nu]}$
is imaginary self-dual $\star F_{\mu\nu}=i F_{\mu\nu}$.  
This is exactly the same form as those 
appearing in extended supergravity models considered thus far.\footnote{
More precisely, this is true except for the gauged supergravity in which the ``trombone symmetry'' 
is gauged.  In this case, $b_{(1)\rho}{}^\rho {}_{[ij]} =0$ is not satisfied~\cite{LeDiffon:2011wt}. 
This accords with the intuition since this kind of gaugings contributes positively to the cosmological constant and even more this theory does not have a covariant action (even in the electric gaugings). 
It would be interesting to understand better this fact and its relation to the positivity conditions. 
} 
It therefore turns out that the conditions~(\ref{positivity})
and (\ref{app_Killing}) are closely related to the construction of extended supergravity. 
Note however that the condition (\ref{app_sol_AB}) is not sufficient to probe that the 
Witten-Nester energy is positive nor the supergravity embedding is possible. 
For example, the connection $\mas A_{\mu i}{}^j$ in the maximal gauged supergravity
is not ${\rm U}(8)$ but ${\rm SU}(8)$ [i.e, ${\rm Tr}(A_\mu)=0$], which corresponds to the R-symmetry.

Since $V^\mu =i \bar \epsilon^i \gamma^\mu \epsilon_i$ generates an asymptotic 
time translation at infinity, the condition (\ref{app_Killing}) 
requires that this asymptotic symmetry is enhanced to the exact symmetry 
for the configuration in which
$\hat \nabla_\mu \epsilon_i=0$ is satisfied. This is in accordance with the 
intuition that the BPS states are in mechanical equilibrium for which gravitational 
attractions and moduli fields are compensated by the electromagnetic 
repulsive forces, implying the existence of the Killing field.

Though we did not discuss the new types of connections in the body of text,
the results of this appendix will be instrumental for 
constructing (bosonic sector of) supergravity incorporating 
noncanonical scalar fields and constraining modified theories of gravity.

\end{document}